\pdfoutput=1

\documentclass[11pt,twoside,a4paper,cmspaper,final,collab]{cms-tdr}
\def\svnVersion{384716}\def\cmsCernNoTag{CERN-EP/2016-224}\def\cmsCernDate{\today}\def\cmsMessage{Published in Physical Review D as \href{http://dx.doi.org/10.1103/PhysRevD.95.01.012009}{\doi{10.1103/PhysRevD.95.01.012009.}}}

\begin{document}\cmsNoteHeader{SUS-14-020}

\hyphenation{had-ron-i-za-tion}
\hyphenation{cal-or-i-me-ter}
\hyphenation{de-vices}
\RCS$Revision: 384682 $
\RCS$HeadURL: svn+ssh://svn.cern.ch/reps/tdr2/papers/SUS-14-020/trunk/SUS-14-020.tex $
\RCS$Id: SUS-14-020.tex 384682 2017-02-02 16:44:36Z jchu $
\newlength\cmsFigWidth
\ifthenelse{\boolean{cms@external}}{\setlength\cmsFigWidth{0.35\textwidth}}{\setlength\cmsFigWidth{0.48\textwidth}}
\ifthenelse{\boolean{cms@external}}{\providecommand{\cmsLeft}{top\xspace}}{\providecommand{\cmsLeft}{left\xspace}}
\ifthenelse{\boolean{cms@external}}{\providecommand{\cmsRight}{bottom\xspace}}{\providecommand{\cmsRight}{right\xspace}}
\ifthenelse{\boolean{cms@external}}{\providecommand{\cmsTable}[1]{#1}}{\providecommand{\cmsTable}[1]{\resizebox{\textwidth}{!}{#1}}}
\providecommand{\cPV}{\ensuremath{\cmsSymbolFace{V}}\xspace}
\providecommand{\HT}{\ensuremath{H_\mathrm{T}}\xspace}
\providecommand{\CLs}{\ensuremath{\mathrm{CL}_\mathrm{s}}\xspace}
\ifthenelse{\boolean{cms@external}}{\providecommand{\NA}{\ensuremath{\cdots}}\xspace}{\providecommand{\NA}{\text{---}}\xspace}
\ifthenelse{\boolean{cms@external}}{\providecommand{\CL}{C.L.\xspace}}{\providecommand{\CL}{CL\xspace}}\newcommand{\taunu}{\ensuremath{c\tau}}
\newcommand{\Mnu}{\ensuremath{M}\xspace}
\newcommand{\clearmu}{\ensuremath{\mu_{\text{clear}}}\xspace}
\newcommand{\clearsig}{\ensuremath{\sigma_{\text{clear}}}\xspace}
\newcommand{\shapepar}{\ensuremath{\boldsymbol\nu}}
\newcommand{\thesig}{\ensuremath{\taunu = 1\mm$, $\Mnu = 400\GeV}\xspace}
\newcommand{\dbv}{\ensuremath{d_{\mathrm{BV}}}\xspace}
\newcommand{\dvv}{\ensuremath{d_{\mathrm{VV}}}}
\newcommand{\dvvc}{\ensuremath{d_{\mathrm{VV}}^{\kern 0.15em\mathrm{C}}}\xspace}
\newcommand{\phiv}{\ensuremath{\phi_{\mathrm{BV}}}\xspace}
\newcommand{\dphivv}{\ensuremath{\Delta\phi_{\textrm{VV}}}\xspace}
\newcommand{\asi}{\ensuremath{a^{(s)}_i}\xspace}
\newcommand{\abi}{\ensuremath{a^{(b)}_i}\xspace}
\newcommand{\Asi}{\ensuremath{A^{(s)}_i}\xspace}
\newcommand{\Abi}{\ensuremath{A^{(b)}_i}\xspace}
\newcommand{\hatAsi}{\ensuremath{\hat{A}^{(s)}_i}\xspace}
\newcommand{\hatAbi}{\ensuremath{\hat{A}^{(b)}_i}\xspace}
\newcommand{\sigbi}{\ensuremath{{\sigma^{(b)}_i}}\xspace}

\cmsNoteHeader{SUS-14-020}
\title{Search for \texorpdfstring{$R$}{R}-parity violating supersymmetry with displaced vertices in proton-proton collisions at \texorpdfstring{$\sqrt{s}=8$\TeV}{sqrt(s)=8 TeV}}

\date{\today}

\abstract{
Results are reported from a search for $R$-parity violating supersymmetry
in proton-proton collision events collected by the CMS experiment
at a center-of-mass energy of $\sqrt{s}=8$\TeV.  The data sample
corresponds to an integrated luminosity of 17.6\fbinv.  This
search assumes a minimal flavor violating model in which the lightest
supersymmetric particle is a long-lived neutralino or gluino,
leading to a signal with jets emanating from displaced vertices.
In a sample of events with two displaced vertices, no excess
yield above the expectation from standard model processes is
observed, and limits are placed on the pair production cross section
as a function of mass and lifetime of the neutralino or gluino.
At 95\% confidence level, the
analysis excludes cross sections above approximately 1\unit{fb} for neutralinos or gluinos with mass between 400 and 1500\GeV
and mean proper decay length between 1 and 30\mm.
Gluino masses are excluded below 1 and 1.3\TeV
for mean proper decay lengths of 300\mum and 1\mm, respectively, and
below 1.4\TeV for the range 2--30\mm.
The results are also applicable to other models in which
long-lived particles decay into multijet final states.
}

\hypersetup{%
pdfauthor={CMS Collaboration},%
pdftitle={Search for R-parity violating supersymmetry with displaced vertices in proton-proton collisions at sqrt(s)=8 TeV},%
pdfsubject={CMS},%
pdfkeywords={CMS, physics, SUSY, RPV, displaced vertices}}

\maketitle

\section{Introduction}

In spite of extensive efforts by the ATLAS and CMS Collaborations at
the CERN LHC, the superpartners of
standard model (SM) particles predicted by supersymmetry
(SUSY)~\cite{Nilles:1983ge,Haber:1984rc} have not yet been observed.
If superpartners are produced and
$R$-parity~\cite{Farrar:1978xj} is conserved, the lightest supersymmetric particle (LSP)
passes through the detector unobserved, except for a potentially large amount of missing
transverse energy.  The assumption of $R$-parity conservation is
motivated by experimental observations such as limits on the proton
lifetime~\cite{Weinberg:1981wj}.  This assumption is not strictly required as long as
either lepton or baryon number is conserved, or the associated
$R$-parity violating (RPV)~\cite{Barbier:2004ez} terms in the
Lagrangian are extremely small.
Searches for a variety of signatures
have not yet found any evidence for RPV SUSY~\cite{gluinoATLAS,Chatrchyan:2013gia,Aad:2015lea,Aad:2014iza,SUS-14-003}.

In minimal flavor violating (MFV) models of RPV
SUSY~\cite{Nikolidakis:2007fc,Yuval}, the Yukawa couplings between
superpartners and SM particles are the sole source of flavor
symmetry violation, and the amplitudes for lepton- and baryon-number
changing interactions are correspondingly small.  At the LHC, the LSP typically
decays within the detector volume, so there is no large missing
transverse energy.  The production processes of the superpartners
are similar to those in the minimal supersymmetric standard
model in that superpartners are produced in pairs, but the
phenomenology depends on the identity of the LSP.

\begin{figure}[hbtp]
\centering
\includegraphics[width=\cmsFigWidth]{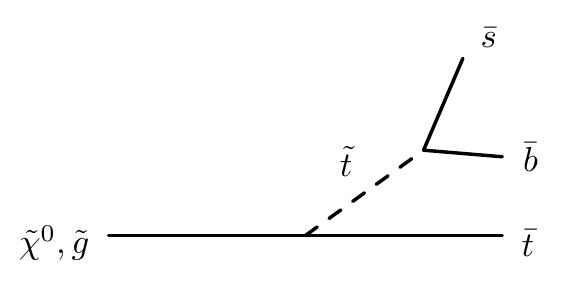}
\caption{
Decay diagram for the pair-produced neutralino (\PSGcz) or gluino (\PSg) LSP in the
assumed signal model. In both cases, the LSP decays into a top
antiquark plus a virtual top squark (\PSQt); the top squark then decays via a
baryon-number violating process into strange and bottom antiquarks.
}
\label{fig:diagram}
\end{figure}

This analysis uses a benchmark signal model described
in Ref.~\cite{Yuval}, in which the LSP is assumed to be either a neutralino
or a gluino that is sufficiently heavy to decay into a
top antiquark and a virtual top squark. The virtual top squark
then decays via a baryon-number violating process to strange and
bottom antiquarks, as shown in Fig.~\ref{fig:diagram}. Although this decay is heavily
suppressed by the Yukawa coupling, it still dominates the top squark
rate, with other partial widths being suppressed by a factor of 100 or
more. As a consequence, the LSP is long-lived, with a lifetime that
depends on the model parameters. For large parts of the parameter
space, pair-produced LSPs lead to interesting signatures. Observable
effects include increased top quark production rates; events with many
jets, especially \PQb quark jets; and events with displaced vertices.

The decay of the LSP results in multiple jets
emerging from a displaced vertex, often with wide opening angles.
To identify the displaced vertices, we use a custom vertex reconstruction algorithm 
optimized for these distinctive features.
This algorithm differs from standard methods used to identify \PQb quark
jets~\cite{BTV-12-001}, which assume a single jet whose momentum is aligned
with the vertex displacement from the primary vertex.  
Our signature consists of two 
vertices, well separated in space.  Studies based on event samples from Monte Carlo (MC) simulation
show that SM background events rarely contain even one such reconstructed
displaced vertex. In the even rarer events with two displaced
vertices, the vertices are usually not well separated from each other.

The CMS Collaboration has also searched for pairs of
displaced jets from a single vertex~\cite{EXO-12-038}, while
this analysis searches for a pair of displaced
vertices, each of which is associated with a jet. The study reported here is sensitive to 
mean proper decay lengths between 300\mum and 30\mm, which are 
shorter than those probed by a similar analysis performed by the ATLAS
Collaboration~\cite{ATLASDV}, and longer than those probed by a CMS analysis
that looked for prompt LSP decays based on the jet and \PQb-tagged jet
multiplicity distributions~\cite{SUS-14-003}.

This analysis applies not only to the MFV model described here, but
more generally to models for physics beyond the SM with long-lived
particles decaying to multiple jets.  In addition to the results of
the search with a neutralino or gluino LSP, we present a method for
reinterpretation of the analysis.

\section{The CMS detector}
\label{sec:detector}

The central feature of the CMS detector is a superconducting
solenoid providing a magnetic field of 3.8\unit{T} aligned with the
proton beam direction.  Contained within the field volume of the
solenoid are a silicon pixel and strip tracker, a lead tungstate
electromagnetic calorimeter (ECAL), and a brass and scintillator
hadronic calorimeter (HCAL).  Outside the solenoid is the steel
magnetic return yoke interspersed with muon tracking chambers.  A
more detailed description of the CMS detector, together with a
definition of the coordinate system used and the relevant kinematic
variables, can be found in Ref.~\cite{Chatrchyan:2008zzk}.

The silicon tracker, which is particularly relevant to this
analysis, measures the tracks of charged particles in the range of
pseudorapidity, $\eta$, up to $\abs{\eta} < 2.5$.  For nonisolated
particles with transverse momentum, \pt, of 1 to 10 \GeV and
$\abs{\eta} < 1.4$, the track resolutions are typically 1.5\% in
\pt, 25--90 \mum in the impact parameter in the transverse plane,
and 45--150 \mum in the impact parameter in the longitudinal
direction~\cite{TRK-11-001}.  When combining information from the
entire detector, the jet energy resolution amounts typically to 15\%
at 10\GeV, 8\% at 100\GeV, and 4\% at 1\TeV, to be compared to about
40\%, 12\%, and 5\% obtained when the ECAL and HCAL calorimeters
alone are used~\cite{Chatrchyan:2011ds}.

The first level (L1) of the CMS trigger system, which is composed of custom
hardware processors, uses information from the calorimeters and muon
detectors to select the most interesting events in a fixed time
interval of less than 4\mus.  The high-level trigger (HLT) processor
farm further decreases the event rate from around 100\unit{kHz} to
less than 1\unit{kHz}, before data storage.

\section{Event samples}
\label{sec:samples}
The data sample used in this analysis corresponds to an integrated
luminosity of 17.6\fbinv, collected in proton-proton (pp) collisions
at a center-of-mass energy of $\sqrt{s}=8\TeV$ in 2012.  Events are
selected using a trigger requiring the presence of at least four
jets reconstructed from energy deposits in the calorimeters.  At the
L1 trigger, the jets are required to have $\pt > 40\GeV$, while in the HLT the threshold is $\pt > 50\GeV$.  The
latter threshold is afforded by a special data-taking strategy
called ``data parking"~\cite{CMS-DP-2012-022}, in which the
triggered events were saved but not promptly reconstructed, allowing
a higher event rate.  The data included in this analysis represent
the fraction of the 2012 LHC operation for which this strategy was
implemented.

Simulated events are used to model both the signal and background
processes.  Using \PYTHIA 8.165~\cite{PYTHIA}, signal samples with
varying neutralino masses $M$ ($200 \leq M \leq 1500$\GeV) and lifetimes $\tau$
($0.1 \leq c\tau \leq 30$\mm) are produced.  In these samples, neutralinos
are produced in pairs; each neutralino is forced to undergo a
three-body decay into top, bottom, and strange \mbox{(anti-)quarks}.
Backgrounds arising from SM processes are dominated by multijet and
top quark pair (\ttbar) events.  The multijet processes include \PQb quark pair events.  Smaller contributions come from single top quark
production (single t), vector boson production in association with
additional jets (\cPV{}+jets), diboson production ($\cPV\cPV$), and top quark
pairs with a radiated vector boson (\ttbar+$\cPV$).  Processes with a
single vector boson include virtual photons, $\PW$ bosons, or \Z bosons,
while the diboson processes include $\PW\PW$, $\PW\Z$, and $\Z\Z$.  Single top
events are simulated with \POWHEG 1.0~\cite{Nason:2004rx,
Frixione:2007vw, Alioli:2010xd, Alioli:2009je, Re:2010bp}; diboson
events are simulated with \PYTHIA 6.426~\cite{PYTHIA6}; all other
backgrounds are simulated using \MADGRAPH 5.1~\cite{MADGRAPH}.  For
all samples, hadronization and showering are done using \PYTHIA
6.426 with tune Z2*.  The Z2* tune is derived from the Z1
tune~\cite{Field:2010bc}, which uses the CTEQ5L parton distribution
set, whereas Z2* adopts CTEQ6L~\cite{Pumplin:2002vw}.  The detector
response for all simulated samples is modeled using a
\GEANTfour-based simulation~\cite{GEANT} of the CMS detector.  The
effects of additional $\Pp\Pp$ interactions per bunch crossing
(``pileup'') are included by overlaying additional simulated
minimum-bias events, such that the resulting distribution of the
number of interactions matches that observed in the experiment.

\section{Event preselection}
\label{sec:eventsel}

To ensure that the four-jet trigger efficiency is high and well
understood, more stringent criteria are applied offline, requiring
at least four jets in the calorimeter with $\pt > 60\GeV$.  These
jets are reconstructed from calorimeter energy deposits, which are
clustered by the anti-\kt
algorithm~\cite{Cacciari:2008gp, Cacciari:2011ma} with a distance
parameter of 0.5.  The trigger efficiency determined using events
satisfying a single-muon trigger is $(96.2 \pm 0.2)$\% for events
with four offline jets with $\pt > 60\GeV$.  The simulation
overestimates this efficiency by a factor of $1.022 \pm 0.002$, so,
where used, its normalization is corrected by this amount.

Jets considered in the rest of the analysis are those obtained in
the full event reconstruction performed using a particle-flow (PF)
algorithm~\cite{CMS-PAS-PFT-09-001,CMS-PAS-PFT-10-001}.  The PF
algorithm reconstructs and identifies photons, electrons, muons, and
charged and neutral hadrons with an optimized combination of
information from the various elements of the CMS detector.  Before
clustering the PF candidates into jets, charged PF candidates are
excluded if they originate from a pp interaction vertex other than
the primary vertex, which is the one with the largest scalar
$\Sigma\abs{\pt}^2$.  The resulting particles are clustered into
jets, again by the anti-\kt algorithm with a distance
parameter of 0.5.  Jets used in the analysis must satisfy $\pt >
20\GeV$ and $\abs{\eta} < 2.5$.

For an event to be selected for further analysis, the scalar sum of
the \pt of jets in the event \HT is required to be at
least 500\GeV. This requirement has little impact on signal events
but is useful for suppressing SM background.

\section{Vertex reconstruction, variables, and selection}
\label{sec:vertexreco}

\subsection{Vertex reconstruction}
Displaced vertices are reconstructed from tracks in the CMS silicon
tracker.  These tracks are required to have $\pt > 1\GeV$, at least
eight measurements in the tracker including one in the pixel
detector, and a transverse impact parameter with respect to the beam
axis of at least 100\mum.  The impact parameter requirement favors
vertices that are displaced from the primary vertex.  The vertex
reconstruction algorithm starts by forming seed vertices from all
pairs of tracks that satisfy these requirements.  Each vertex is
fitted with the Kalman filter approach~\cite{Kalman}, and a fit is
considered successful if it has a $\chi^2$ per degree of freedom
($\chi^2$/dof) that is less than 5.  The vertices are then merged
iteratively until no pair of vertices shares tracks.  Specifically,
for each pair of vertices that shares one or more tracks, if the
three-dimensional (3D) distance between the vertices is less than
4 times the uncertainty in that distance, a vertex is fit to the
tracks from both, and they are replaced by the merged vertex if the
fit has $\chi^2/\mathrm{dof} < 5$.  Otherwise, each track is assigned to
one vertex or the other depending on its 3D impact parameter
significance with respect to each of the vertices, as follows:

\begin{itemize}
\item if the track is consistent with both vertices (both impact
parameters less than 1.5 standard deviations), assign it to the
vertex that has more tracks already;
\item if the track's impact parameter is greater than 5 standard
deviations from either vertex, drop it from that vertex;
\item otherwise, assign the track to the vertex to which it has a
smaller impact parameter significance.
\end{itemize}

Each remaining vertex is then refit, and if the fit satisfies the
requirement of $\chi^2/\mathrm{dof} < 5$, the old vertex is replaced
with the new one; otherwise it is dropped entirely.

This algorithm is similar in many regards to those used to identify
(``tag") \PQb quark jets~\cite{BTV-12-001}.  Typical \PQb tagging
algorithms, however, are optimized for identifying the decay in
flight of a particle into a single jet and consequently make
requirements that degrade sensitivity to the multijet final states
sought here.  For example, \PQb tagging algorithms generally require
that the tracks assigned to a vertex are approximately aligned with
the flight direction from the primary vertex to the decay point,
which is inefficient when there are multiple jets in the final
state, including some that may be directed at large angles with
respect to the flight path.  The \PQb tagging algorithms also discard
tracks with impact parameters beyond those typical for \PQb quark
daughters (${>}2\mm$), thereby significantly reducing the efficiency
for finding vertices with large displacements.

\subsection{Vertex variables and selection}
\label{sec:vertexvarssel}
The vertexing procedure produces multiple vertices per event, only
some of which are consistent with the signal.  In order to select
quality vertices, we impose additional requirements on the vertex
and its associated tracks and jets.  The requirements for each
vertex are:

\begin{itemize}
\item{at least five tracks;}
\item{at least three tracks with $\pt > 3\GeV$;}
\item{at least one pair of tracks with separation $\Delta R < 0.4$,
where $\Delta R = \sqrt{\smash[b]{(\Delta\eta)^2 + (\Delta\phi)^2}}$,
to favor vertices that include multiple tracks from a single jet;}
\item{at least one pair of tracks with $\Delta R > 1.2$ to favor
vertices involving multiple jets;}
\item{$\Delta R < 4$ for all pairs of tracks, to suppress wide-angle
track coincidences;}
\item{at least one jet that shares one or more tracks with the
vertex;}
\item{displacement in $x$-$y$ of the vertex from the detector origin
of less than 25\mm, to suppress vertices from interactions in the
beam pipe or detector material;}
\item{uncertainty in the $x$-$y$ distance of the vertex from the
beam axis of less than 25\mum.}
\end{itemize}

In the data, 181\,076 events have one vertex satisfying the above
requirements, 251 have two of them, and no events have more than
two.  The candidate sample is composed of two-vertex events.

\subsection{Signal discrimination in two-vertex events}
The signal is extracted from the two-vertex events using the spatial
separation between the vertices.  In signal events, the two LSPs are
emitted approximately back-to-back, leading to large separations.
We define the distance between the two vertices in the $x$-$y$ plane
as $\dvv$, and fit this distribution to extract the signal.  The fit
to the observed $\dvv$ distribution is described in
Sec.~\ref{sec:statinterp}.

\begin{figure}[tbp!]
\centering
\includegraphics[width=0.48\textwidth]{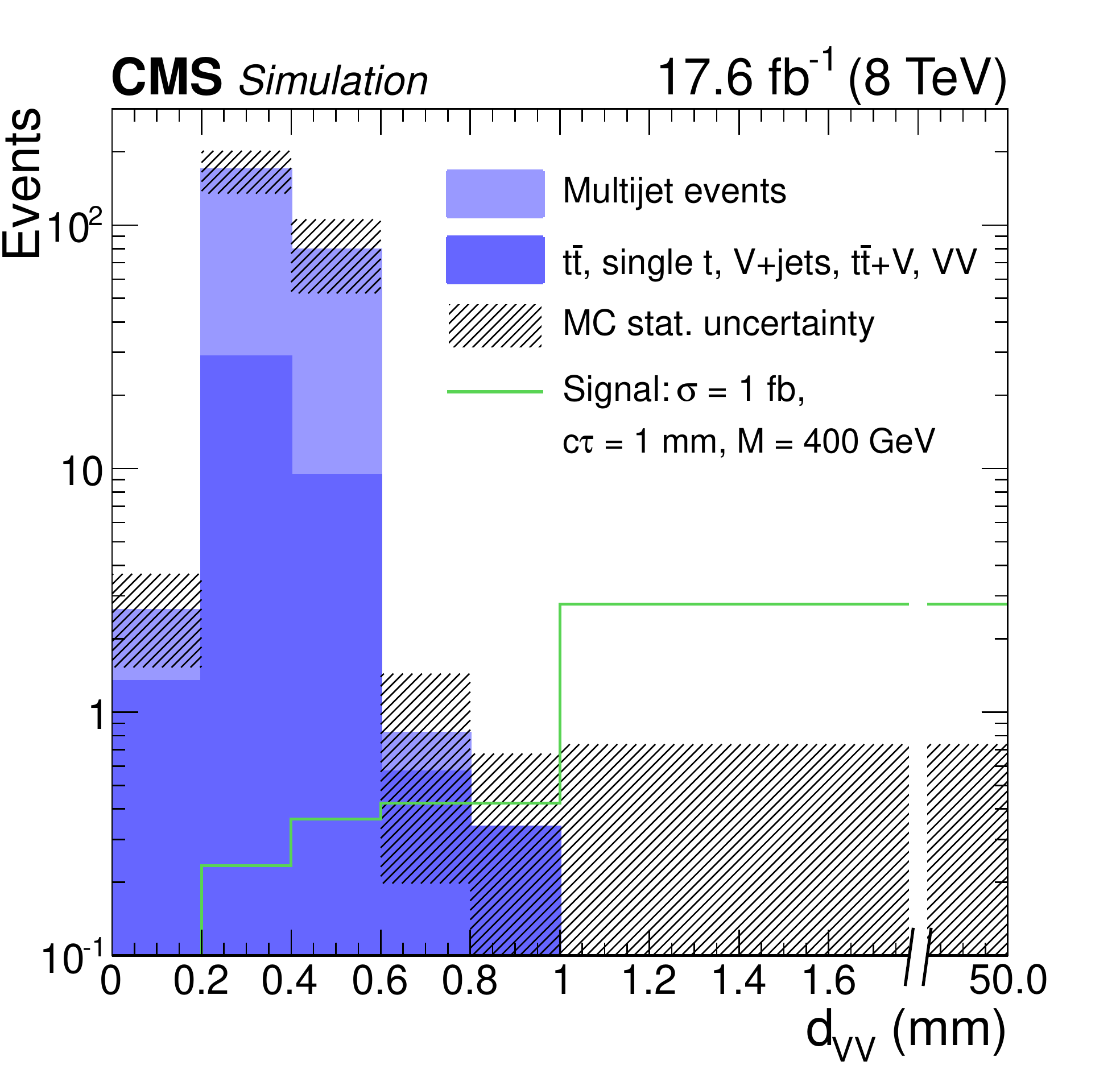}
\caption{Distribution of the $x$-$y$ distance between vertices, $\dvv$, for a
simulated signal with LSP $\taunu = 1\mm$, $M = 400\GeV$, and production
cross section 1\unit{fb}, overlaid on simulated background
normalized to the observed number of events.  All vertex and event
selection criteria have been applied.  In the last bin where there 
are no simulated background events, the shaded band represents an 
approximate 68\% confidence level upper limit.
}
\label{fig:MCsvdist2d}
\end{figure}

The signal $\dvv$ templates are taken directly from simulation, with
a distinct template for each LSP mass $M$ and lifetime $\tau$.  In signal
simulation, fewer than 10\% of events in the candidate sample have
more than two selected vertices.  For these events, the two vertices
with the highest number of tracks are selected for the $\dvv$
calculation, and in the case where two vertices have the same number
of tracks, the vertex with decay products that have the higher
invariant mass is chosen.  The mass is reconstructed using the
momenta of the associated tracks, assuming that the particles
associated with the tracks have the charged pion mass.
Figure~\ref{fig:MCsvdist2d} shows the $\dvv$ distribution of an
example simulated signal with $\taunu = 1\mm$, $M = 400\GeV$, and
production cross section 1\unit{fb}, overlaid on the simulated
background.  The bins in $\dvv$ are chosen to be sensitive to the
peaking nature of the background at low $\dvv$; five 200\micron bins
are used from 0 to 1\mm, then one bin from 1 to 50\mm where the
contribution from the long-lived signal dominates.

\begin{figure}[tbp!]
\centering
\includegraphics[width=0.48\textwidth]{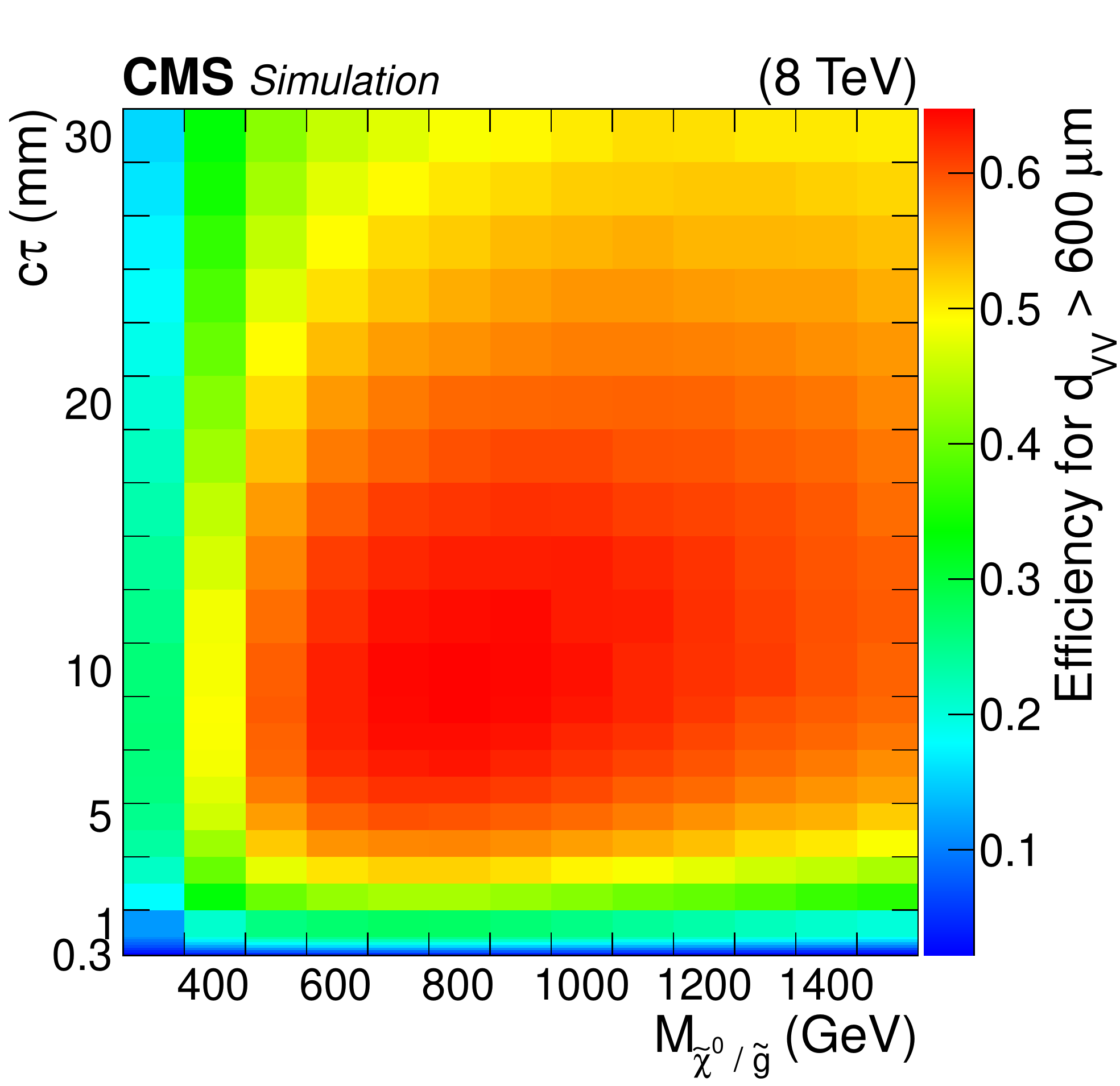}
\caption{Signal efficiency as a function of neutralino or gluino mass
and lifetime. All vertex and event selection criteria have been applied, as well as the requirement $\dvv > 600\mum$.}
\label{fig:sigeff}
\end{figure}

Figure~\ref{fig:sigeff} shows the signal efficiency as a function of LSP mass and lifetime 
in the region $\dvv > 600\mum$, where the background is low.  The signal
efficiency generally increases as lifetime increases, until the lifetime is so long that
decays more often occur beyond our fiducial limit at the beam pipe.  The
efficiency also generally increases as mass increases, up to approximately
800\GeV where it begins to decrease because of the event
selection criteria, particularly the limit on the
opening angle between track pairs in a vertex.

\section{Background template}
\label{sec:bkgest}
Background vertices arise from poorly measured tracks.  These tracks
can arise from the same jet, or from several jets in multijet events.
Because it is an effect of
misreconstruction, two-vertex background events are the coincidence
of single background vertices.

Multijet events and \ttbar production contribute 85\% and 15\% of
the background in the two-vertex sample, respectively.  Other
sources of background, such as V+jets and single t events, are
negligible.  Approximately half of the background events include one
or more \PQb quark jets, whose displaced decay daughters combine with
misreconstructed tracks to form vertices.

Instead of relying on simulation to reproduce the background, we
construct a background template, denoted by $\dvvc$, from data.
Taking advantage of the fact that two-vertex background events can
be modeled using the one-vertex events, we define a control sample
that consists of the 181\,076 events with exactly one vertex.  Each
value entering the $\dvvc$ template is the distance in the $x$-$y$
plane between two toy vertices, each determined by a value of the
$x$-$y$ distance from the beam axis to the vertex, denoted by
$\dbv$, and a value of the azimuthal angle of the vertex, denoted by
$\phiv$.

\begin{figure}[tbp!]
\centering
\includegraphics[width=0.48\textwidth]{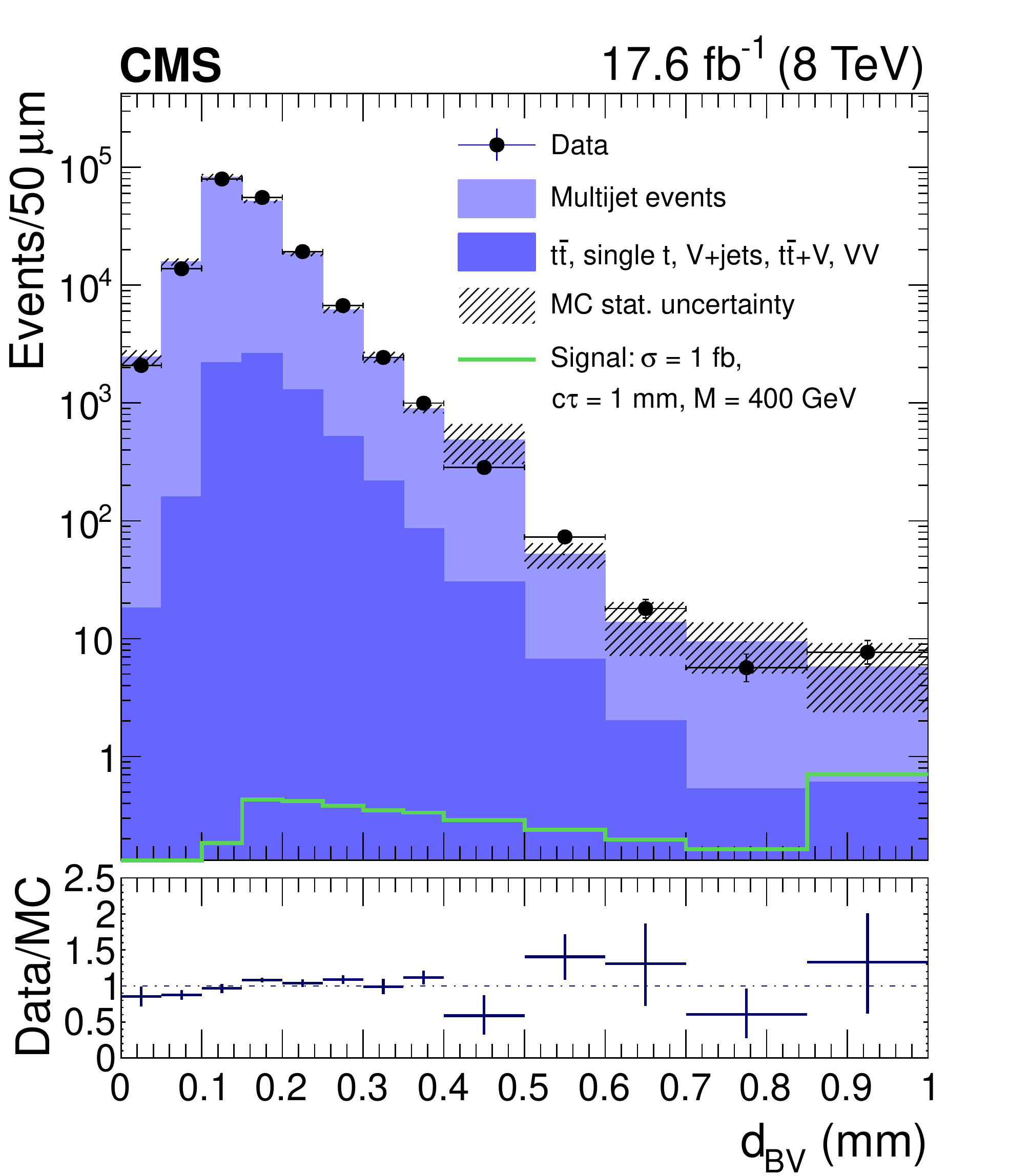}
\caption{One-vertex events: distribution of the $x$-$y$ distance
from the beam axis to the vertex, $\dbv$, for data, simulated
background normalized to data, and a simulated signal with LSP
$\taunu = 1\mm$, $M = 400\GeV$, and production cross section
1\unit{fb}.  Event preselection and vertex selection criteria have
been applied.  The last bin includes the overflow events.}
\label{fig:bs2ddist}
\end{figure}

The two values of $\dbv$ are sampled from the distribution of $\dbv$
for the one-vertex sample, which is shown in
Fig.~\ref{fig:bs2ddist}.  The observed distribution is in good
agreement with the sum of the background contributions from
simulation.

The two values of $\phiv$ are chosen using information about the jet
directions in a one-vertex event.  Since background vertices come
from misreconstructed tracks, they tend to be located perpendicular
to jet momenta.  Therefore, we select a jet at random, preferring
those with larger \pt because of their higher track multiplicity,
and sample a value of $\phiv$ from a Gaussian distribution with
width 0.4 radians, centered on a direction perpendicular to the jet
in the transverse plane. To obtain the second value of $\phiv$, we
repeat this procedure using the same one-vertex event, allowing the
same jet to be chosen twice.

The vertex reconstruction algorithm merges neighboring vertices.  To
emulate this behavior in our background template construction
procedure, we discard pairs of vertices that are not sufficiently
separated.  We keep pairs of vertices with a probability
parametrized by a Gaussian error function with mean $\clearmu$ and
width $\clearsig$.  The values of $\clearmu$ and $\clearsig$, which
are related to the position uncertainties of the tracks, are varied
in the fit to the observed $\dvv$ distribution.  The values found in
the fit are $\clearmu = 320\mum$ and $\clearsig = 110\mum$.

\begin{figure}[tbp!]
\centering
\includegraphics[width=0.48\textwidth]{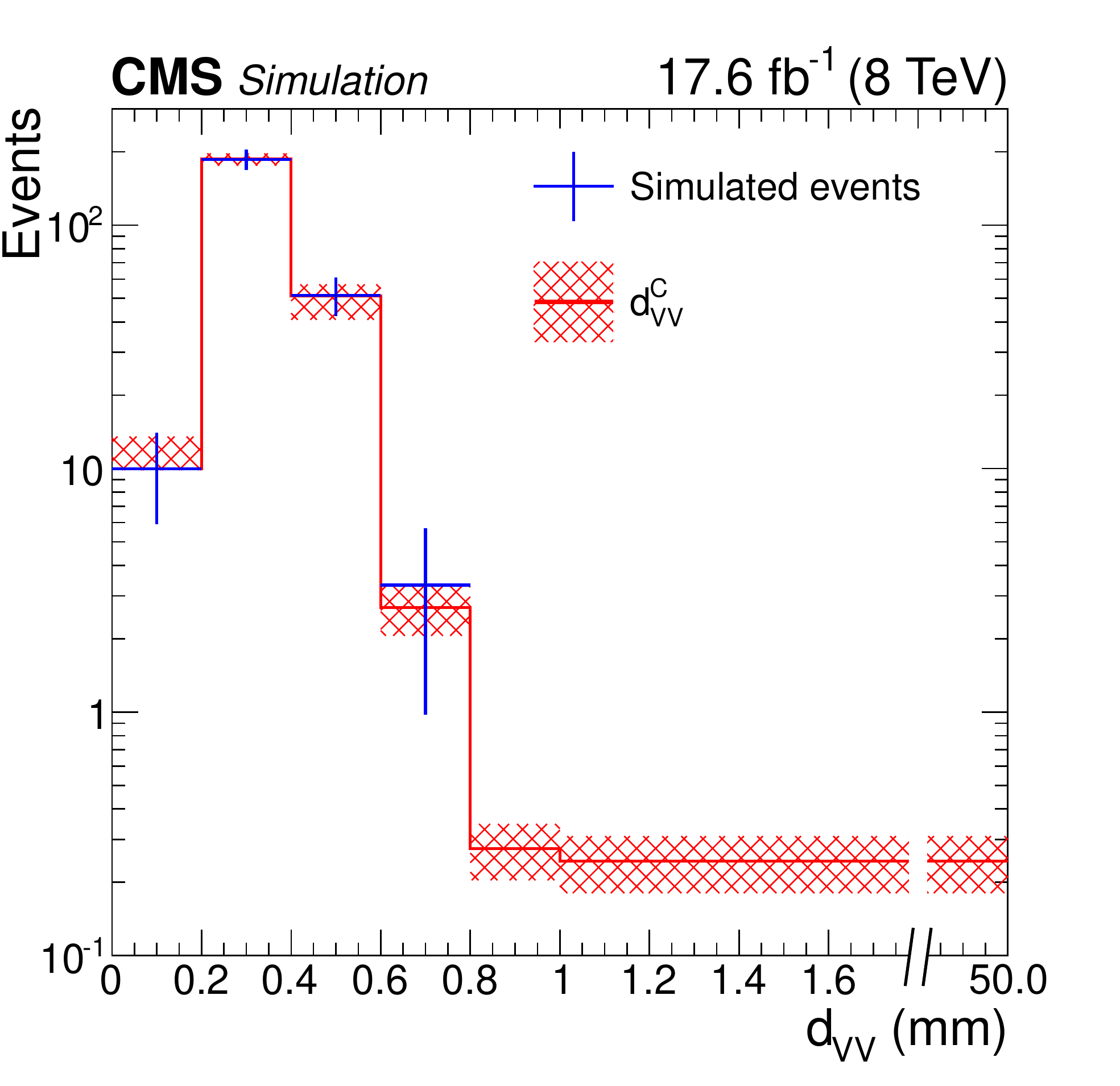}
\caption{Distribution of the $x$-$y$ distance between vertices, $\dvv$, for
simulated background events (blue crosses), overlaid on the template
$\dvvc$ (red line and crosshatches) constructed from simulated one-vertex events.  The distributions are
normalized to the observed number of two-vertex events.  The error
bars for the simulated events represent only the statistical
uncertainty, while the shaded region for the template
is the result of varying $\clearmu$ and $\clearsig$
within one standard deviation of the values from the fit.}
\label{fig:closure}
\end{figure}

Figure~\ref{fig:closure} compares the $\dvvc$ and
$\dvv$ distributions in simulated events, and shows the variation in
$\dvvc$ for values of $\clearmu$ and $\clearsig$ that
are within one standard deviation of the fit values.
The agreement
is well
within the statistical uncertainty.  When normalized to the observed
number of two-vertex events, the difference in their yields in the
region $\dvv > 600\mum$ is $0.6\pm2.6$ events.

\section{Systematic uncertainties}
\label{sec:systematics}

The signal is extracted from a fit of a weighted sum of the signal
and background templates to the observed $\dvv$ distribution.  For
the signal, the simulation provides both the $\dvv$ distribution and
its normalization, and systematic uncertainties arise from sources
such as vertex reconstruction efficiency, track reconstruction,
track multiplicity, pileup conditions, the detector alignment, and
the jet energies.  For the background, for which the template is
derived from a control sample, the systematic uncertainties come
from effects that could cause a discrepancy between the
constructed $\dvvc$ distribution and the nominal $\dvv$ distribution.

\subsection{Systematic uncertainties related to signal distribution and efficiency}
\label{sec:signalsyst}

The dominant systematic uncertainty in the signal normalization
arises from the difference between the vertexing efficiencies in the
simulation and data.  This effect is evaluated in an independent study in
which artificial signal-like vertices are produced in background
events by displacing tracks associated with jets by a known
displacement vector, and then applying the vertex reconstruction algorithm.  The
magnitude of the displacement vector is sampled from an exponential distribution
with scale parameter 1\mm, restricted to values between 0.3 and
25\mm, similar to the expected distribution of signal vertices.  The
direction is calculated from the momentum of the jets in the event,
but is smeared to emulate the difference between the flight and
momentum directions in simulated signal events due to track
inefficiency and unaccounted neutral particles.  Events are required
to satisfy the preselection requirements described in
Sec.~\ref{sec:eventsel}, and the displaced jets satisfy $\pt >
50\GeV$ and $\Delta R < 4$ for all pairs.  To estimate the vertexing
efficiency, we evaluate the fraction of events in which a
vertex satisfying the requirements described in
Sec.~\ref{sec:vertexvarssel} is reconstructed within 50\mum of
the artificial vertex.

This fraction is evaluated for
different numbers of displaced light parton or \PQb quark jets, with the
ratio of efficiencies between data and simulation approaching unity
for larger numbers of jets, independent of the size of the
displacement.
The largest disagreement between data and simulation occurs for the
case where tracks from two light parton jets are displaced, where
the fraction is 70\% in simulation and 64\% in data, with
negligible statistical uncertainty.
The ratio of efficiencies between data and simulation gives an 8.6\%
uncertainty per vertex.
For two-vertex events, the uncertainty is 17\%.

Additional studies explore the sensitivity of other effects that
could alter the signal template.  The vertex clustering depends on
the number of charged particles in the event, which can vary
based on the model of the underlying
event used in \PYTHIA~\cite{CMS-GEN-14-001}.  The signal
templates resulting from the choice of the underlying event model differ by no more than 1\% in any bin and the overall
efficiency changes by no more than 3\%.  This 3\% is taken as a
systematic uncertainty.

To test the sensitivity to a possible misalignment, the signal samples
have been reconstructed using several tracker misalignment scenarios
corresponding to various ``weak modes'': coherent distortions of the
tracker geometry left over by the alignment procedure that lead to a
systematic bias in the track parameters for no penalty in $\chi^2$
of the overall alignment fit~\cite{weakmodes}.  These misalignments
change the overall efficiency by no more than 2\%, which is
taken as a systematic uncertainty.

To study sensitivity to the pileup distribution, we vary the
inelastic $\Pp\Pp$ cross section used in the pileup weighting by
$\pm$5\%~\cite{CMS-FWD-11-001}.  This variation is found to have an
effect of less than 1\% on the signal efficiency.

The uncertainty in the jet energy scale affects the total energy
measured, and could change whether an event
passes the jet \pt or \HT selections.  This
effect is studied by varying the jet energy scale and
resolution~\cite{Chatrchyan:2011ds}, and is found to change the
signal efficiency by less than 1\%.  A 2.6\%
uncertainty~\cite{CMS-PAS-LUM-13-001} is associated with the
integrated luminosity for the 2012 data set and the derived signal
cross section. The uncertainty in the trigger efficiency is less
than 1\%.

Table \ref{tab:sig_systematics} summarizes the systematic
uncertainties in the signal efficiency.  We assume there are no
correlations among them, so we add them in quadrature to obtain the
overall uncertainty.

\begin{table}[tbp!]
\centering
\topcaption{Summary of systematic uncertainties in the signal efficiency.}
\begin{scotch}{lc}
Systematic effect           & Uncertainty (\%) \\
\hline
Vertex reconstruction       & 17   \\
Underlying event            & 3   \\
Tracker misalignment        & 2   \\
Pileup                      & 1 \\
Jet energy scale/resolution & 1 \\
Integrated luminosity       & 3 \\
Trigger efficiency          & 1   \\
\hline
Overall                     & 18 \\
\end{scotch}
\label{tab:sig_systematics}
\end{table}

\subsection{Systematic uncertainties related to background estimate}

The $\dvvc$ background template is constructed from a large sample
of events with a single vertex.  Systematic uncertainties in the
$\dvvc$ template are estimated by varying the $\dvvc$ construction
method and taking the difference between the $\dvvc$ distributions
using the default and alternate methods.  The method for
constructing $\dvvc$ involves drawing two values of $\dbv$ and two
values of $\phiv$, with an angle between vertices $\dphivv$, so the
main uncertainties come from effects related to the $\dbv$ and
$\dphivv$ distributions.

The production of \PQb quarks in pairs introduces a correlation between
the vertex distances in two-vertex events that is not accounted for
when single vertices are paired at random.  In simulation, events
without \PQb quarks have a mean $\dbv$ of ${\sim}160\mum$, while events
with \PQb quarks, which account for 15\% of one-vertex events, have a
mean $\dbv$ of ${\sim}190\mum$, without significant dependence on \PQb quark momentum.   We quantify this effect by sorting
the simulated background events into those with and without \PQb quarks, constructing the $\dvvc$ distributions for each, and then
combining them in the proportions 45:55, which is the ratio of
\PQb-quark to non-\PQb-quark events in two-vertex background events 
determined from simulation.  The systematic uncertainty is taken to
be the difference between the simulated yields obtained with this procedure
and the standard one, scaled to the observed two-vertex yield.

The $\dvvc$ construction method discards pairs of vertices that
would overlap, consistently leading to a two-vertex angular
distribution that peaks at ${\pm}\pi$ radians.  To assess the
systematic uncertainty related to assumptions about the angular
distribution between vertices, we draw $\dphivv$ from the angular
distribution between vertices in simulated two-vertex background
events.  This leads to a $\dvvc$ distribution with a more strongly
peaked $\dphivv$ distribution, and provides a conservative estimate
of the uncertainty.

The statistical uncertainty from the limited number of one-vertex
events that are used to construct the two-vertex distribution is
studied using a resampling method.  Using the $\dbv$ distribution as
the parent, we randomly sample ten new $\dbv$ pseudodata
distributions, and use each to construct a $\dvvc$ distribution.
The root-mean-square variation in bin-by-bin yields in the set of
distributions gives the statistical uncertainty.

There is a small contribution to the uncertainty in the prediction
of $\dvvc$ due to the binning of the $\dbv$ parent distribution;
moving the $\dbv$ tail bin edges around by an amount compatible with
the vertex position resolution, 20\mum, varies the prediction in
$\dvvc$ only in the last two bins: by 0.06 events in the
0.8--1.0\mm bin, and by 0.09 events in the 1.0--50\mm bin.

The results of these four studies are summarized in
Table~\ref{tab:bkgestsyst}.  In assessing the overall systematic
uncertainty in the background template, we add in quadrature the
values and their uncertainties, assuming no correlations.

\begin{table*}[tbp!]
\centering
\topcaption{Systematic uncertainties in the background yield in each
$\dvv$ bin arising from construction of the $\dvvc$ template.  In the first two rows, shifts are given with their
statistical uncertainty.  The last row gives the overall systematic
uncertainties, assuming no correlations.  All yields are normalized
to the observed total number of two-vertex events.}
\cmsTable{
\begin{scotch}{rrrrrrr}
\multirow{2}{*}{Systematic effect}   & \multicolumn{6}{c}{$\dvv$ range} \\ \cline{2-7}
 & \multicolumn{1}{c}{0.0--0.2\mm} & \multicolumn{1}{c}{0.2--0.4\mm} & \multicolumn{1}{c}{0.4--0.6\mm} & \multicolumn{1}{c}{0.6--0.8\mm} & \multicolumn{1}{c}{0.8--1.0\mm} & \multicolumn{1}{c}{1.0--50\mm} \\
\hline
$\dbv$ correlations & $-0.65 \pm 0.05$     & $-3.60 \pm 1.01$     & $ 3.59 \pm 0.76$     & $ 0.63 \pm 0.18$     & $ 0.01 \pm 0.07$     & $ 0.01 \pm 0.04$     \\
$\dphivv$ modeling  & $ 0.74 \pm 0.01$     & $ 1.00 \pm 0.00$     & $ 1.04 \pm 0.01$     & $ 1.18 \pm 0.07$     & $-0.01 \pm 0.06$     & $-0.01 \pm 0.04$     \\
$\dbv$ sample size  & $ 0.05$              & $ 0.54$              & $ 0.51$              & $ 0.17$              & $ 0.04$              & $ 0.07$              \\
$\dbv$ binning      & \NA                    & \NA                    & \NA                    & \NA                    & $ 0.06$              & $ 0.09$              \\
\hline
Overall             & $ 1.0 $              & $ 3.9 $              & $ 3.8 $              & $ 1.4 $              & $ 0.1 $              & $ 0.1 $              \\
\end{scotch}
\label{tab:bkgestsyst}
}
\end{table*}

In principle, there can also be uncertainties in the background template due to the effects
described in Sec.~\ref{sec:signalsyst}.
To assess the impact of the
underlying event and possible tracker misalignment, we generate five million all-hadronic $\ttbar$
events for each scenario, but observe no change in $\dvvc$
larger than 1\%. In addition, we
vary the inelastic pp cross section used in pileup weighting by
$\pm5\%$, the number of pileup interactions, and the jet
energy scale and resolution, and observe effects at the
percent-level or less in each case.  Since the normalization of the
template is a free parameter of the fit, uncertainties such as those
in the integrated luminosity, trigger efficiency, and vertex
reconstruction efficiency do not enter.

\section{Fitting, signal extraction, and statistical interpretation}
\label{sec:statinterp}

The distribution of $\dvv$, the separation between vertices in the
$x$-$y$ plane for two-vertex events, is used to discriminate between
signal and background, with the signal templates taken directly from
the MC simulation and the background template constructed from the
observed one-vertex event sample.  In the following sections, we describe the
fitting and statistical procedures used for the search.

\subsection{Fitting procedure}

To estimate the signal and background event yields,
a binned shape fit is performed using an extended maximum
likelihood method.  Initially neglecting terms arising from
uncertainty in the templates, the log-likelihood function is given by

\begin{equation}
\label{eqn:lnLsimple}
\log \mathcal{L}(n_i | s, b, \shapepar) = \sum_i \left[ n_i \log a_i\left(s, b, \shapepar\right) - a_i\left(s, b, \shapepar\right) \right].
\end{equation}
Here $n_i$ is the number of observed events in bin $i$; $s$ and $b$
are the normalizations of the signal and background templates
corresponding to the yields; $\shapepar$ denotes the shape
parameters $\clearmu$ and $\clearsig$ used in the background
template construction procedure, as described in
Sec.~\ref{sec:bkgest}; and

\begin{equation}
\label{eqn:abkgsig}
a_i\left(s, b, \shapepar\right) = s \asi + b \abi(\shapepar)
\end{equation}
is the weighted sum of the signal and background frequencies $\asi$ and
$\abi$ in bin $i$.

The only assumed shape uncertainty in the signal templates is that
due to the finite MC statistics; the uncertainty is as high as 20\%
for the lowest lifetime and mass samples, but is generally no more
than 1\% in any bin for the majority of the templates.  For the
background templates, a Gaussian uncertainty is assumed in the value
of the template in each bin, truncated at zero.  To incorporate
these uncertainties in the signal and background templates, a
procedure similar to that of Barlow and Beeston~\cite{barlowbeeston}
is followed, modified to allow a bin-by-bin Gaussian uncertainty in
the background shape~\cite{Conway-PhyStat}.  The final log-likelihood
function is then given by

\begin{multline}
\label{eqn:lnL}
\log \mathcal{L}(n_i | s, b, \shapepar, \Asi, \Abi) =  \sum_i n_i \log A_i - A_i \\
                                             + \sum_i M \asi \log M \Asi - M \Asi \\
                                             + \sum_i - \frac{1}{2} {\left( \frac{\abi - \Abi}{\sigbi} \right)}^2,
\end{multline}
with $A_i = s \Asi + b \Abi$.  The $\Asi$ and $\Abi$ replace the
$\asi$ and $\abi$ from above in the shape fit to the data, and are
allowed to vary as either Poisson ($\Asi$) or Gaussian ($\Abi$)
distributed parameters. The quantity $M$ is the number of events
from the MC signal sample that produced the $\asi$ estimates, and
$\sigbi$ are the widths of the Gaussian distributions taken to be the relative
sizes of the uncertainties listed in Table~\ref{tab:bkgestsyst}.
The modified Barlow-Beeston procedure finds the $\Asi$ and $\Abi$
that maximize $\log \mathcal{L}$ given $(s,b,\shapepar)$; the
difference here is that the $\Abi$ are Gaussian distributed
parameters.

The likelihood function is only weakly dependent on the background
shape parameters $\shapepar$, and when signal is injected, the best
fit values $\hat\shapepar$ agree well with the background-only
values.  The fit is well behaved: for most signal templates, in
pseudo-experiments where the true signal and background strengths
are known, the distribution of the fitted yields for $s$ and $b$
have means consistent with those input, and the widths of the
distributions as measured by their root-mean-square are consistent
with the uncertainties in the fits.  For the signal templates with
low lifetimes, however, the signal yield is biased downward when an
injected signal is present.  This is due to the background shape
being allowed to vary upward at high $\dvv$ within the uncertainties
assigned.  When no injected signal is present, there is a
bias toward obtaining $s > 0$ when fitting using templates with
$\taunu < 300\mum$.  Therefore, we only consider signals with
$\taunu \geq 300\mum$ in the fit and the search.

\subsection{Statistical analysis}
The test statistic $q$ used to quantify any excess of signal events
over the expected background is given by a profile likelihood
ratio~\cite{Rolke2005493}:

\begin{equation}
\label{eqn:teststat}
q = \log \frac{\max_{s \geq 0, b \geq 0} \mathcal{L}(n_i | s, b, \hat{\shapepar}, \hatAsi, \hatAbi)} {\max_{b \geq 0} \mathcal{L}(n_i | s = 0, b, \hat{\shapepar}, \hatAsi, \hatAbi)},
\end{equation}
where for each value of $s$ and $b$ the nuisance parameters
$\hatAsi$, $\hatAbi$, and $\hat{\shapepar}$ are found that maximize
the relevant likelihood.  The probability under the background-only
hypothesis, $p_0$, to obtain a value of the test statistic at least
as large as that observed, $q_\text{obs}$, is estimated as the
fraction of 10\,000 pseudo-experiments with $q \geq q_\text{obs}$.
This is referred to as the $p$-value for a particular signal
hypothesis.  The pseudo-experiments are generated using the
background $\dvvc$ distribution corresponding to the background-only
$\hat{\shapepar}$, and background count $b$ drawn from a Poisson
distribution with mean equal to $n$, the number of events in the
data.  The nuisance parameters $\shapepar$, $\Asi$, and $\Abi$ are
drawn from their corresponding Poisson or Gaussian distributions in
each pseudo-experiment.

We obtain limits on the signal yield, which can be converted into
limits on the product of the cross section for neutralino or gluino
pair production and the square of the branching fraction for decay
via the channel under study, denoted by $\sigma\mathcal{B}^2$.  To
obtain limits on $\sigma\mathcal{B}^2$, for a given number of signal
events $s_0$, we calculate the probability for the null hypothesis of
$s=s_0$ versus the alternative that $s<s_0$ denoted by $p_{s_0}$.
We do this in practice by generating 10\,000 pseudo-experiments with
$s$ drawn from a Poisson distribution with mean $s_0$, and $b$ drawn
from a Poisson distribution with mean $n - s_0$.  The background
shape $\dvvc$ is taken from the $\shapepar$ from the original fit
and signal shape corresponding to the signal hypothesis in question,
with $\Abi$ from their Gaussian distributions.  The null hypothesis
probability $p_{s_0}$ is then the fraction of pseudo-experiments
where $q \geq q(s_0)$.  We protect against downward fluctuations in
the data by using the
\CLs criterion~\cite{CLS1,CLS2}, defining the
statistic as

\begin{equation}
\label{eqn:CLs}
\CLs = \frac{p_{s_0}}{1 - p_0}.
\end{equation}
The 95\% confidence level (\CL) upper limit on $s$ is then the
biggest $s_0$ for which \CLs is still greater
than 0.05.

The limit on the signal yield is converted to a limit on
$\sigma\mathcal{B}^2$ using the efficiencies calculated from
simulation and the integrated luminosity of the data sample,
17.6\fbinv.  We include the effect of the estimated 18\% signal
efficiency uncertainty by varying the cross section in each
pseudo-experiment by the value sampled from a log-normal density
with location parameter 1 and scale parameter 0.18.

\subsection{Results of the fit}

\begin{figure}[tbp!]
\centering
\includegraphics[width=0.48\textwidth]{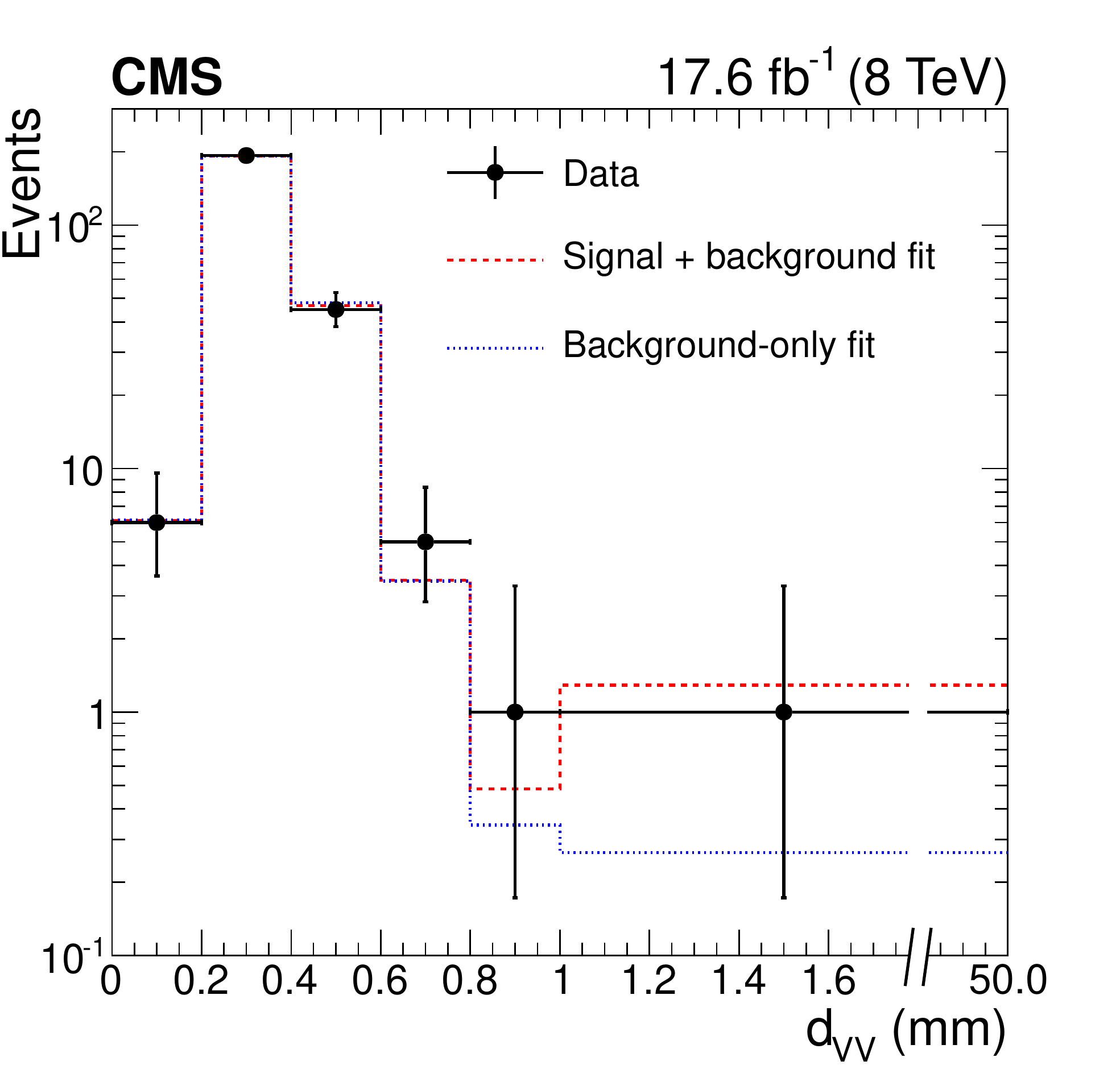}
\caption{ The observed distribution of the $x$-$y$ distance between
  the vertices, $\dvv$, shown as points with error
  bars. Superimposed are the results of the fits with the
  background-only (blue dotted lines) and signal+background (red
  dashed lines) hypotheses, using the signal template corresponding to LSP \thesig.}
\label{fig:datafit}
\end{figure}

The result of the fit to data is shown in Fig.~\ref{fig:datafit},
for the LSP \thesig signal template. The observed counts in each bin,
along with the predictions from the background-only fit and the
related uncertainties, are listed in Table~\ref{tab:bincounts}.
There is a small excess of events with $0.6 < \dvv < 50\mm$: 7 in
the data, while the background-only fit predicts $4.1 \pm 1.4$,
where the uncertainty is the overall systematic uncertainty
discussed in Sec.~\ref{sec:systematics}.  In the
signal+background fits, a typical value for the signal yield is $1.7
\pm 1.9$, obtained with the \thesig signal hypothesis.  The
associated $p$-value obtained from pseudo-experiments is in the
range 0.05--0.14 for signals with $0.3 \leq \taunu \leq 30\mm$, with
the larger $p$-values coming from those with longer lifetimes.

\begin{table}[tbp!]
\centering
\topcaption{Observed and expected background event yields in
each bin.  The uncertainty is the sum in quadrature of the
statistical and systematic uncertainties.}
\begin{scotch}{llrr}
Bin $i$ & $\dvv$ range & Observed $n_i$ & Expected event yield \\
\hline
1       & 0.0--0.2\mm  &   6 & $  6.2 \pm 1.0$ \\
2       & 0.2--0.4\mm  & 193 & $192.6 \pm 3.9$ \\
3       & 0.4--0.6\mm  &  45 & $ 48.1 \pm 3.8$ \\
4       & 0.6--0.8\mm  &   5 & $  3.5 \pm 1.4$ \\
5       & 0.8--1.0\mm  &   1 & $  0.3 \pm 0.1$ \\
6       & 1.0--50\mm   &   1 & $  0.3 \pm 0.1$ \\
\end{scotch}
\label{tab:bincounts}
\end{table}

\subsection{Upper limits on signal cross section}

\begin{figure}[tbp!]
\centering
\includegraphics[width=0.48\textwidth]{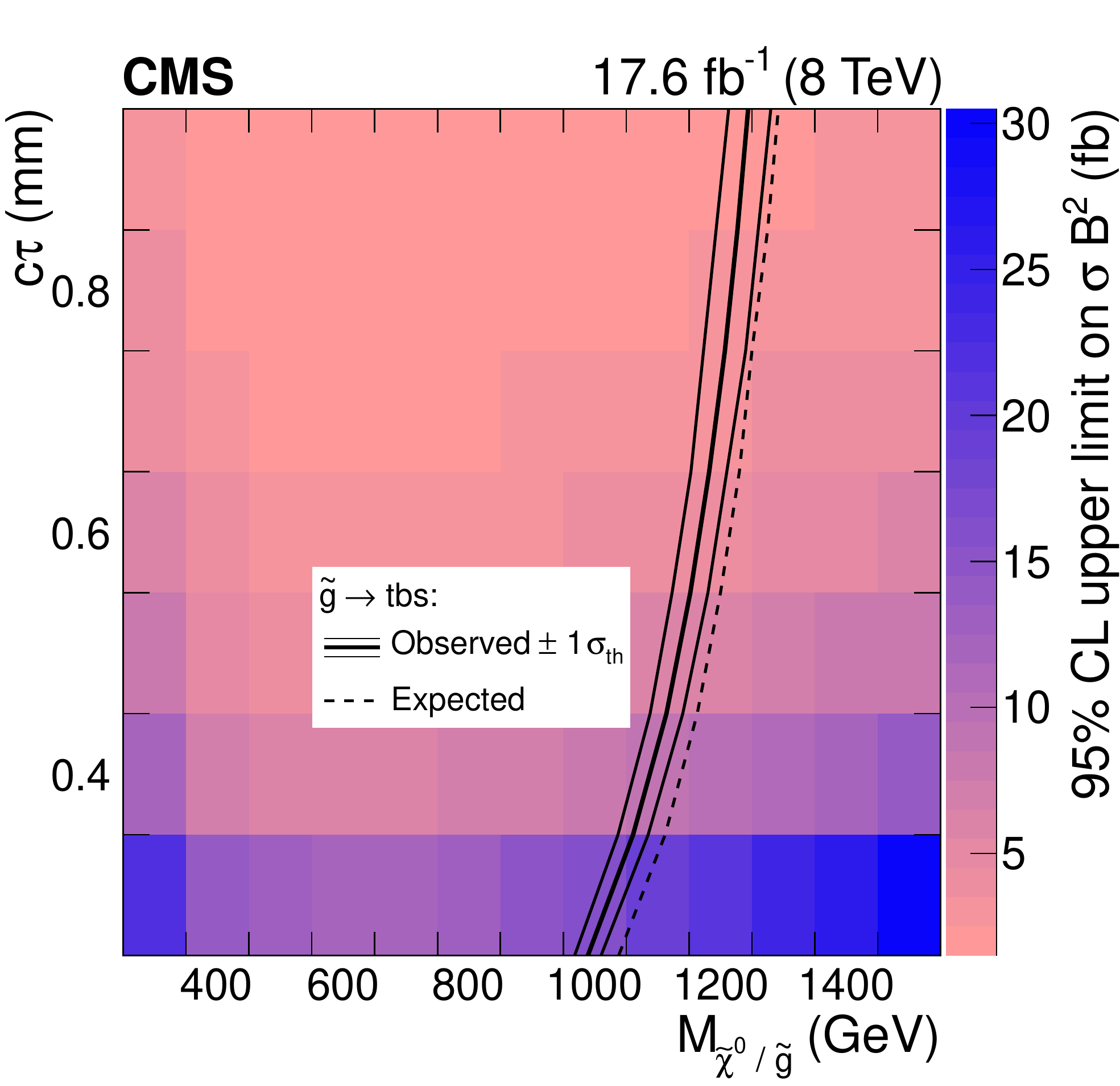}
\includegraphics[width=0.48\textwidth]{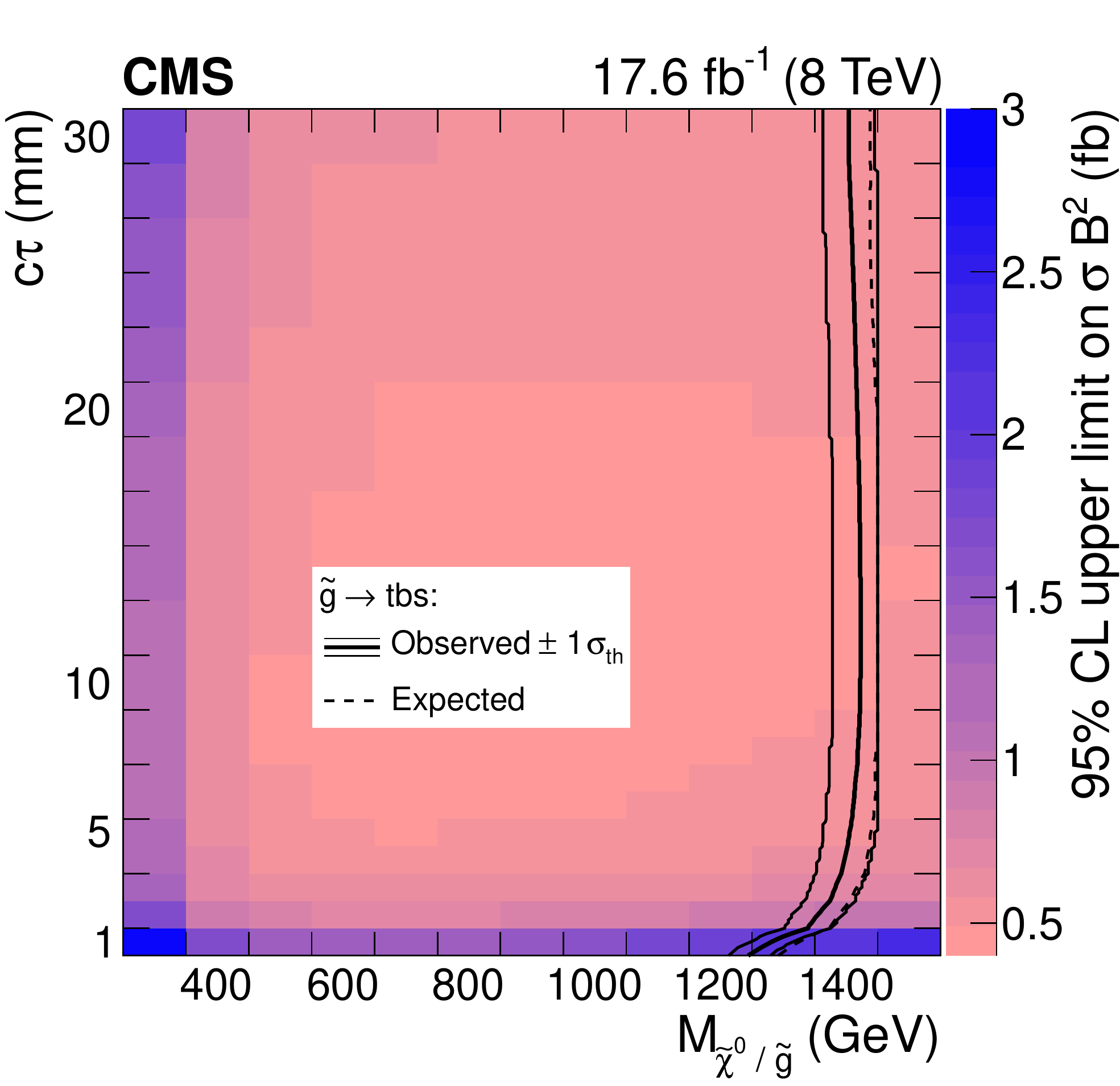}
\caption{Observed 95\% CL upper limits on cross section times
branching fraction squared, with overlaid curves assuming gluino
pair production cross section, for both observed (solid), with
$\pm$1 standard deviation theoretical uncertainties, and expected
(dashed) limits.  The search excludes masses to the left of the
curve.  The \cmsLeft plot spans $\taunu$ from 300 through 900\mum, while
the \cmsRight plot ranges from 1 to 30\mm.}
\label{fig:datalimits}
\end{figure}

Figure~\ref{fig:datalimits} shows the observed 95\% CL upper limits
on $\sigma\mathcal{B}^2$.  As an example, for a neutralino with mass
of 400\GeV and $\taunu$ of 10\mm, the observed 95\% CL upper limit
on $\sigma\mathcal{B}^2$ is 0.6\unit{fb}.

Exclusion curves are overlaid, assuming the gluino pair production
cross section~\cite{Beenakker:1996ch,Kulesza:2008jb,Kulesza:2009kq,Beenakker:2009ha,Beenakker:2011fu}.
In the context of the MFV model that we are studying, either a neutralino or a gluino LSP can decay into the final state targeted in the search.

The scan in $\taunu$ is in steps of 100\mum from 300\mum to 1\mm,
then in 1\mm steps up to 10\mm, and in 2\mm steps to 30\mm; the mass
points are spaced by 100\GeV.  The exclusion curves are produced by
linear interpolation of the limit scan, which identifies the set of points
for which the interpolated upper limit is less than the gluino pair
production cross section (the neutralino pair production cross
section is expected to be much smaller).

\section{Extending the search to other signal models}
\label{sec:theoryinterp}

The search for displaced vertices applies to other types of
long-lived particles decaying to multiple jets.  Here we present a
generator-level selection that can be used to reinterpret the
results of our analysis.  For signal models in which there are two
well-separated displaced vertices, this generator-level selection
approximately replicates the reconstruction-level efficiency.  The
selection is based on the displacements of the long-lived particles,
and the momenta and angular distributions of their daughter
particles, which are taken to be \PQu, \PQd, \PQs, \PQc, and \PQb quarks;
electrons; and muons.  The daughter particles are said to be
``accepted" if they satisfy $\pt > 20\GeV$ and $\abs{\eta} < 2.5$, and
``displaced" if their transverse impact parameter with respect to
the origin is at least 100\mum.  The criteria of the
generator-level selection are:
\begin{itemize}
\item{at least four accepted quarks with $\pt > 60\GeV$;}
\item{\HT of accepted quarks $> 500\GeV$;}
\item{for each vertex:}
\begin{itemize}
\item{$x$-$y$ distance from beam axis $<$25\mm;}
\item{at least one pair of accepted displaced daughter particles with $\Delta R > 1.2$;}
\item{$\Delta R < 4$ for all pairs of accepted displaced daughter particles;}
\item{at least one accepted displaced daughter quark;}
\item{$\sum{\pt}$ of accepted displaced daughter particles $> 200\GeV$;}
\end{itemize}
\item{$x$-$y$ distance between vertices $> 600\mum$.}
\end{itemize}

In the region with $\dvv > 600\mum$, the background level is well
determined and is insensitive to fit parameters.  Use of this
generator-level selection replicates the reconstruction-level
efficiency with an accuracy of 20\% or better for a selection of
models for which the signal efficiency is high (${>}10\%$).  The
selection may underestimate the trigger efficiency because it does
not take into account effects such as initial- and final-state
radiation, and may overestimate the efficiency for reconstructing
vertices with \PQb quark secondaries, since the \PQb quark lifetime can
impede the association of their decay products with the reconstructed
vertices.

\section{Summary}
\label{sec:conclusion}

A search for $R$-parity violating SUSY in which long-lived neutralinos or gluinos
decay into multijet final states
was performed using proton-proton collision events collected with the CMS
detector at $\sqrt{s}=8\TeV$ in 2012.
The data sample corresponded to
an integrated luminosity of 17.6\fbinv, and was collected
requiring the presence of at least four jets.  No
excess above the prediction from standard model processes was observed, and
at 95\%
confidence level, the data excluded cross section times branching fraction squared above
approximately 1\unit{fb} for neutralinos or gluinos with mass
between 400 and 1500\GeV and $\taunu$ between 1 and 30\mm.
Assuming gluino pair production cross sections,
gluino masses below 1 and 1.3\TeV were excluded
for mean proper decay lengths of 300\mum and 1\mm, respectively, and
below 1.4\TeV for the range 2--30\mm.
While
the search specifically addressed $R$-parity violating SUSY, the
results were relevant to other massive particles that decay to two or
more jets.  These are the most restrictive bounds to date on the production and
decay of pairs of such massive particles with intermediate lifetimes.

\begin{acknowledgments}
We congratulate our colleagues in the CERN accelerator departments for the excellent performance of the LHC and thank the technical and administrative staffs at CERN and at other CMS institutes for their contributions to the success of the CMS effort. In addition, we gratefully acknowledge the computing centers and personnel of the Worldwide LHC Computing Grid for delivering so effectively the computing infrastructure essential to our analyses. Finally, we acknowledge the enduring support for the construction and operation of the LHC and the CMS detector provided by the following funding agencies: BMWFW and FWF (Austria); FNRS and FWO (Belgium); CNPq, CAPES, FAPERJ, and FAPESP (Brazil); MES (Bulgaria); CERN; CAS, MoST, and NSFC (China); COLCIENCIAS (Colombia); MSES and CSF (Croatia); RPF (Cyprus); SENESCYT (Ecuador); MoER, ERC IUT, and ERDF (Estonia); Academy of Finland, MEC, and HIP (Finland); CEA and CNRS/IN2P3 (France); BMBF, DFG, and HGF (Germany); GSRT (Greece); OTKA and NIH (Hungary); DAE and DST (India); IPM (Iran); SFI (Ireland); INFN (Italy); MSIP and NRF (Republic of Korea); LAS (Lithuania); MOE and UM (Malaysia); BUAP, CINVESTAV, CONACYT, LNS, SEP, and UASLP-FAI (Mexico); MBIE (New Zealand); PAEC (Pakistan); MSHE and NSC (Poland); FCT (Portugal); JINR (Dubna); MON, RosAtom, RAS, and RFBR (Russia); MESTD (Serbia); SEIDI and CPAN (Spain); Swiss Funding Agencies (Switzerland); MST (Taipei); ThEPCenter, IPST, STAR, and NSTDA (Thailand); TUBITAK and TAEK (Turkey); NASU and SFFR (Ukraine); STFC (United Kingdom); DOE and NSF (USA).

\hyphenation{Rachada-pisek} Individuals have received support from the Marie-Curie program and the European Research Council and EPLANET (European Union); the Leventis Foundation; the A. P. Sloan Foundation; the Alexander von Humboldt Foundation; the Belgian Federal Science Policy Office; the Fonds pour la Formation \`a la Recherche dans l'Industrie et dans l'Agriculture (FRIA-Belgium); the Agentschap voor Innovatie door Wetenschap en Technologie (IWT-Belgium); the Ministry of Education, Youth and Sports (MEYS) of the Czech Republic; the Council of Science and Industrial Research, India; the HOMING PLUS program of the Foundation for Polish Science, cofinanced from European Union, Regional Development Fund, the Mobility Plus program of the Ministry of Science and Higher Education, the National Science Center (Poland), contracts Harmonia 2014/14/M/ST2/00428, Opus 2013/11/B/ST2/04202, 2014/13/B/ST2/02543 and 2014/15/B/ST2/03998, Sonata-bis 2012/07/E/ST2/01406; the Thalis and Aristeia programs cofinanced by EU-ESF and the Greek NSRF; the National Priorities Research Program by Qatar National Research Fund; the Programa Clar\'in-COFUND del Principado de Asturias; the Rachadapisek Sompot Fund for Postdoctoral Fellowship, Chulalongkorn University and the Chulalongkorn Academic into Its 2nd Century Project Advancement Project (Thailand); and the Welch Foundation, contract C-1845.
\end{acknowledgments}
\bibliography{auto_generated}

\cleardoublepage \appendix\section{The CMS Collaboration \label{app:collab}}\begin{sloppypar}\hyphenpenalty=5000\widowpenalty=500\clubpenalty=5000\textbf{Yerevan Physics Institute,  Yerevan,  Armenia}\\*[0pt]
V.~Khachatryan, A.M.~Sirunyan, A.~Tumasyan
\vskip\cmsinstskip
\textbf{Institut f\"{u}r Hochenergiephysik,  Wien,  Austria}\\*[0pt]
W.~Adam, E.~Asilar, T.~Bergauer, J.~Brandstetter, E.~Brondolin, M.~Dragicevic, J.~Er\"{o}, M.~Flechl, M.~Friedl, R.~Fr\"{u}hwirth\cmsAuthorMark{1}, V.M.~Ghete, C.~Hartl, N.~H\"{o}rmann, J.~Hrubec, M.~Jeitler\cmsAuthorMark{1}, A.~K\"{o}nig, I.~Kr\"{a}tschmer, D.~Liko, T.~Matsushita, I.~Mikulec, D.~Rabady, N.~Rad, B.~Rahbaran, H.~Rohringer, J.~Schieck\cmsAuthorMark{1}, J.~Strauss, W.~Treberer-Treberspurg, W.~Waltenberger, C.-E.~Wulz\cmsAuthorMark{1}
% ...
\vskip\cmsinstskip
% ...
1:~~Also at Vienna University of Technology, Vienna, Austria\\
% ...

\end{sloppypar}
\end{document}